\documentclass[twocolumn,showpacs,preprintnumbers,
superscriptaddress,nofootinbib,10pt,aps,prd,reprint]{revtex4-1}
\usepackage{amsmath,amssymb,bm,fancybox,graphicx,hyperref}
\usepackage[table]{xcolor}
\usepackage[utf8]{inputenc}
\usepackage[T1]{fontenc} 
\definecolor{linkblue}{rgb}{0.,0.,0.9333}
\hypersetup{
    pdffitwindow=true,      
    pdfnewwindow=true,      
    bookmarksnumbered=true,
    colorlinks=true,       
    linkcolor=red,          
    citecolor=darkgreen,        
    filecolor=magenta,      
    urlcolor=cyan,           
    menucolor=blue,
}
\definecolor{darkgreen}{rgb}{0,.7,0}
\definecolor{darkorange}{rgb}{9.,.4,0}
\begin{document}
\title{Anisotropic pressure induced by finite-size effects 
in SU(3) Yang-Mills theory}
\author{Masakiyo Kitazawa}
\affiliation{Department of Physics, Osaka University,
Toyonaka, Osaka 560-0043, Japan}
\affiliation{
J-PARC Branch, KEK Theory Center, 
Institute of Particle and Nuclear Studies, KEK, 
203-1, Shirakata, Tokai, Ibaraki, 319-1106, Japan}
\author{Sylvain Mogliacci}
\affiliation{Department of Physics, University of Cape Town, 
Rondebosch 7701, South Africa}
\author{Isobel Kolb\'e}
\affiliation{Department of Physics, University of Cape Town, 
Rondebosch 7701, South Africa}
\author{W. A. Horowitz}
\affiliation{Department of Physics, University of Cape Town, 
Rondebosch 7701, South Africa}
\begin{abstract}
We study the pressure anisotropy in anisotropic finite-size systems in SU(3) Yang-Mills theory at nonzero temperature. Lattice simulations are performed on lattices with anisotropic spatial volumes with periodic boundary conditions. The energy-momentum tensor defined through the gradient flow is used for the analysis of the stress tensor on the lattice. We find that a clear finite-size effect in the pressure anisotropy is observed only at a significantly shorter spatial extent compared with the free scalar theory, even when accounting for a rather large mass in the latter.
\end{abstract}
\preprint{J-PARC-TH-158}
\maketitle
\section{Introduction}
\label{sec:intro}

Thermodynamic quantities such as the pressure and energy density are
fundamental observables for investigating a thermal medium.
In Quantum Chromodynamics (QCD) and pure Yang-Mills (YM) theories,
the analysis of thermodynamics in first-principle numerical
simulations on the lattice has been performed actively, and
successful results have been established~\cite{Karsch:1982ve,
Boyd:1996bx,Umeda:2008bd,Ejiri:2009hq,Borsanyi:2012ve,Giusti:2012yj,
Asakawa:2013laa,Borsanyi:2013bia,Bazavov:2014pvz,Caselle:2016wsw,
Shirogane:2016zbf,Taniguchi:2016ofw,Kitazawa:2016dsl,Giusti:2016iqr,
Caselle:2018kap,Iritani:2018idk}.
These results have played a crucial role in revealing properties 
of the thermal medium described by these theories,
such as the onset of a deconfinement phase transition.
They also play a critical role in phenomenological studies on 
the dynamics of relativistic heavy-ion
collisions. 

Thermodynamic quantities are usually defined in the thermodynamic limit,
i.e.\ the infinite volume limit, which conventionally refers to an isotropic system which is asymptotically large in all three spatial directions. 
In this limit, the pressure is isotropic due to rotational symmetry.
The stress tensor $\sigma_{ij}$,
which is related to the spatial components of the energy-momentum tensor (EMT)
$T_{\mu\nu}$ as $\sigma_{ij}=-T_{ij}$ ($i,j=1,2,3$),
is then given by
\begin{align}
  \sigma_{ij} = -P \delta_{ij},
  \label{eq:sigma=Pdel}
\end{align}
with pressure $P$.
As the force per unit area acting on a surface with the unit normal
vector $n_j$ is given by $F_i/S=\sigma_{ij} n_j$~\cite{Landau:1982dva},
Eq.~(\ref{eq:sigma=Pdel}) means that the pressure is isotropic and
always perpendicular to the surface.
On the other hand, in a thermal system with a finite volume,
rotational symmetry is broken due to the boundary conditions 
and this effect can give rise to a deviation of the stress tensor from
the form in Eq.~(\ref{eq:sigma=Pdel}).

A well-known example of such a pressure anisotropy is the Casimir
effect~\cite{Casimir:1948dh};
see for reviews~\cite{Ambjorn:1981xw,Plunien:1986ca,Bordag:2009zzd}.
When two perfectly conducting plates are placed within a sufficiently short distance,
there appears an attractive force between the plates due to quantum effects.
This means that the pressure along the direction perpendicular to
the plates becomes negative.
At the spatial points inside the plates 
$\sigma_{ij}$ is no longer proportional to the unit matrix;
$\sigma_{ij}$ has a positive eigenvalue
with the eigenvector perpendicular to the plates, 
while the other two eigenvalues are negative~\cite{Brown:1969na}.
Such an anisotropic structure of $\sigma_{ij}$ is known to survive
even at nonzero temperature~\cite{Ambjorn:1981xw,Plunien:1986ca,
Brown:1969na,Bordag:2009zzd,Mogliacci:2018oea}.

Recently, the numerical simulations for the Casimir effect in YM theory
have been performed for 2+1 dimension~\cite{Karabali:2018ael} and
SU(2) gauge theory~\cite{Chernodub:2018aix}.
In the present study we investigate Casimir-type effects
in the 3+1 dimensional SU(3) YM theory focusing on the anisotropy of
the stress tensor in lattice numerical simulations.

Phenomenologically, the goal of relativistic heavy ion collisions is to connect experimental measurements to verify fundamental knowledge of QCD.
The success of the hydrodynamic models for describing the experimental data
measured at the Relativistic Heavy Ion Collider (RHIC) and the Large
Hadron Collider (LHC)~\cite{Song:2010mg,Gale:2013da,Weller:2017tsr}
implies that these experiments generate the hottest matter in the universe~\cite{Cleymans:2005xv}
with a viscosity to entropy density ratio $\eta/s\sim2/4\pi$~\cite{Song:2010mg,Gale:2013da} close to the conjectured lowest bound~\cite{Kovtun:2004de}.
A fundamental input into these hydrodynamics simulations is
the equation of state (EoS), which is the thermodynamic energy as a function of pressure $\varepsilon(P)$.  Lattice calculations on isotropic lattices extrapolated to the thermodynamic limit have so far provided the most realistic EoS used in these hydrodynamics calculations~\cite{Huovinen:2009yb,Song:2010mg,Gale:2013da,Bluhm:2013yga,Weller:2017tsr}.  More recently, hydrodynamic models have shown remarkable agreement with particle distributions measured in small system collisions~\cite{Bzdak:2013zma,Weller:2017tsr}.  There have also been recent advances in hydrodynamic theory to systems with large pressure anisotropies~\cite{Bazow:2013ifa}.  This recent research begs for an investigation into the QCD EoS in finite-sized, anisotropic systems.
Jet tomography is another important avenue of research in heavy ion collision phenomenology~\cite{Majumder:2010qh,Burke:2013yra}.
While hydrodynamic studies of high multiplicity small system collisions suggest that small droplets of quark-gluon plasma (QGP) are generated in these collisions, high momentum particles do not appear to appreciably lose energy in these small collision systems~\cite{Kolbe:2015rvk}.  It is therefore interesting to investigate the small system corrections to energy loss models based on perturbative QCD (pQCD) methods~\cite{Majumder:2010qh,Horowitz:2012cf,Burke:2013yra},
especially the transverse gluon self-energy and its relation to the Debye screening scale of QCD~\cite{Djordjevic:2003qk}.

In the present study,
in order to investigate a manifestation of the pressure anisotropy
in SU(3) YM theory at nonzero temperature
we measure thermal expectation values of the EMT on lattices with
an anisotropic spatial volume with
periodic boundary conditions (PBC).
To carry out this analysis, we use the so-called gradient flow
method~\cite{Suzuki:2013gza,Kitazawa:2016dsl}.
In this method, thermodynamic quantities are obtained from 
the thermal expectation values of the EMT~\cite{Suzuki:2013gza}
defined through the gradient flow~\cite{Luscher:2010iy,Narayanan:2006rf,Luscher:2011bx}.
The direct determination of the anisotropic stress tensor can indeed be
performed with this method.
We note that other methods for the measurement of thermodynamic quantities on
the lattice, see e.g. Refs.~\cite{Karsch:1982ve,Boyd:1996bx,Umeda:2008bd,
Giusti:2016iqr,Caselle:2018kap}, cannot deal with the anisotropic
stress tensor because they rely on thermodynamic relations valid only
in the infinite and isotropic volume limit\footnote{In SU(3) YM theory, there is an excellent agreement on 
numerous thermodynamic quantities computed using various lattice methods in the limit of infinite and isotropic volume~\cite{Caselle:2018kap,Iritani:2018idk}.}.

We perform numerical simulations on the lattice above the critical temperature $T_c$.
One spatial extent, $L_x$, is set to be shorter than the others, and the effect of the chosen spatial boundary condition
on pressure anisotropy is studied.
The result is compared with the anisotropic pressure in the free massless
and massive scalar field theories.
We find that the effect of the periodic spatial boundary in SU(3) YM theory is
remarkably weaker compared to the one in a free scalar theory, i.e. that a manifestation of the anisotropy in the stress
tensor occurs at significantly smaller $L_xT$.

This paper is organized as follows.
In the next section we summarize basic properties of the EMT in an anisotropic thermal system.
We then introduce the EMT operator on the lattice in Sec.~\ref{sec:EMT}.
After describing the setup of our numerical simulations in 
Sec.~\ref{sec:setup}, we discuss numerical results in Sec.~\ref{sec:result}.
The last section is devoted to discussions and outlook.

\section{Anisotropic pressure}
\label{sec:aniso}

In this section, we summarize basic properties of the EMT
in anisotropic thermal systems.

Throughout this paper, we consider three-dimensional finite-size systems
with PBC along all spatial directions at nonzero temperature $T$.
We further suppose that the spatial extent along the $y$ and $z$ directions
is sufficiently long, $L_y,~ L_z \gg 1/T$, and discuss the response
of the system with respect to the size along the $x$ direction, $L_x$.

In the Matsubara formalism, a system at nonzero temperature 
is described by a field theory in Euclidean four-dimensional space
where the temporal extent $L_\tau=1/T$ with PBC imposed
for bosonic fields.
We denote the EMT in Euclidean space as
$T_{\mu\nu}^{\rm E}(x)$ with $\mu,\nu=1,2,3,4$.
Its thermal expectation value $\langle T_{\mu\nu}^{\rm E}(x) \rangle$
is related to those in Minkowski space
$T_{\mu\nu}(x) $ with $\mu,\nu=0,1,2,3$ as
\begin{align}
  \langle T_{00} \rangle = - \langle T_{44}^{\rm E} \rangle, \quad
  \langle T_{ij} \rangle = \langle T_{ij}^{\rm E} \rangle,
\end{align}
for $i,j=1,2,3$.
The vacuum expectation value of the EMT at $T=0$ is normalized
to vanish, $\langle T_{\mu\nu} \rangle_0=0$.
The energy density is given by $\varepsilon = \langle T_{00} \rangle =
- \langle T_{44}^{\rm E} \rangle$.

When all spatial lengths are sufficiently large, 
$L_x,~L_y,~L_z\gg1/T$, the system obviously has
an approximate rotational symmetry,
and $\langle T_{\mu\nu} \rangle$ is diagonal with spatial components given by
\begin{align}
  \langle T_{ij} \rangle = P \delta_{ij},
\end{align}
where $P$ is the pressure in an isotropic thermal system.
When $L_x\ne L_y$ or $L_z$, the rotational symmetry is broken due to the
boundary conditions.
From the reflection symmetries along individual axes, 
$\langle T_{\mu\nu} \rangle$ is diagonal even in this case\footnote{
  Choosing to rotate the coordinate system outside of the $y-z$ symmetric plane would break this reflection symmetry, and the resulting spatial components of $\langle T_{\mu\nu} \rangle$ would no longer be diagonal.
  } with
\begin{align}
  \langle T_{\mu\nu} \rangle = {\rm diag}( \varepsilon , P_x , P_z , P_z ),
\end{align}
where $P_x=\langle T_{11} \rangle $ and
$P_z= \langle T_{22} \rangle = \langle T_{33} \rangle $ are
the stress along longitudinal and transverse directions.
$\langle T_{22} \rangle = \langle T_{33} \rangle$
due to the rotational symmetry in the $y$-$z$ plane. 

For $L_x=L_\tau=1/T$, the $\tau$ and $x$ directions become symmetric
in the Euclidean space and one obtains
$\langle T_{11} \rangle=\langle T_{11}^{\rm E} \rangle
=\langle T_{44}^{\rm E} \rangle=-\langle T_{00} \rangle$, or
\begin{align}
  P_x=-\varepsilon \qquad (\mbox{recall only for } L_x=1/T).
  \label{eq:px=e}
\end{align}
By writing the trace of the EMT as\footnote{
  We employ the metric $g^{\mu\nu}={\rm diag}(1,-1,-1,-1)$
  in the Minkowski space.}
\begin{align}
  \Delta= \sum_{\mu=0}^3 \langle T_\mu^\mu \rangle
  = -\sum_\mu \langle T_{\mu\mu}^{\rm E} \rangle = \varepsilon - P_x -2P_z,
\end{align}
Eq.~(\ref{eq:px=e}) shows that in this case
\begin{align}
  P_x = -P_z - \frac12 \Delta \qquad (\mbox{for } L_x=1/T).
  \label{eq:xzD}
\end{align}
In particular, when the theory has conformal symmetry
one has 
$\Delta=0$ and $P_x/P_z=-1$ for $L_x=1/T$.
We will see below that the quantum breaking of conformal symmetry
in SU(3) YM theory will yield $P_x/P_z\ne-1$ for $L_x=1/T$.

As PBC are imposed for all directions in the Euclidean space,
the role of the axes can be exchanged.
For example, a Euclidean system of hypervolume
$L_\tau \times L_x \times L_y \times L_z$ can be interpreted
in two different ways~\cite{Brown:1969na}:
\\
(A) Volume $L_x \times L_y \times L_z$ at temperature $T=1/L_\tau$;
\\
(B) Volume $L_\tau \times L_y \times L_z$ at temperature $T=1/L_x$.
\\
In (A) and (B), the role of the components of the EMT is
also exchanged.
The energy density for (A) and (B) is given by 
$\varepsilon=-\langle T_{44}^{\rm E}\rangle$ and 
$\varepsilon=-\langle T_{11}^{\rm E}\rangle$, respectively.
Also, the spatial component of the EMT for (B) is given by
${\rm diag} ( T_{44}^{\rm E} , T_{22}^{\rm E} , T_{33}^{\rm E} )$.

In order to see this explicitly,
let us consider a system at $T=0$ 
with finite $L_x$.
With an infinitesimal variation of $L_x$ given by $d L_x$,
the energy per unit area in the $y$-$z$ plane increases as
$-( \partial L_x \langle T_{44}^{\rm E} \rangle_{L_x} )/( \partial L_x ) d L_x$,
where $ \langle T_{\mu\nu}^{\rm E} \rangle_{L_x}$ is the expectation
value of $T_{\mu\nu}^{\rm E}$ at the length $L_x$.
According to the principle of virtual work, this change is related to
$P_x$ as
\begin{align}
  \langle T_{11}^{\rm E} \rangle_{L_x} = P_x
  = \frac{\partial}{\partial {L_x}} (L_x \langle T_{44}^{\rm E} \rangle_{L_x}).
  \label{eq:virtualwork}
\end{align}
Next, by exchanging the roles of the $\tau$ and $x$ axes in the Euclidean space,
this system can be regarded as a nonzero temperature system with $T=1/L_x$.
By relabeling subscripts of EMT in accordance with the exchange of axes,
Eq.~(\ref{eq:virtualwork}) reads
\begin{align}
  \langle T_{44}^{\rm E} \rangle = \frac{\partial}{\partial(1/T)} \frac{\langle T_{11}^{\rm E} \rangle}T
  \label{eq:G-H}
\end{align}
which is nothing but the Gibbs-Helmholtz relation
\begin{align}
  \varepsilon = - \frac{\partial}{\partial(1/T)} \frac PT,
\end{align}
where we substituted $\langle T_{11}^{\rm E} \rangle=P$ 
because three spatial directions are infinitely large 
in the exchanged coordinates.

In the following numerical analyses, we constrain our attention
to the case $L_x\ge 1/T$.
These results can also be regarded as the system with $L_x\le 1/T$
by exchanging the $\tau$ and $x$ axes.

\section{Energy-momentum tensor on the lattice}
\label{sec:EMT}

In this study we measure the components of the EMT
on the lattice with the use of the EMT operator defined through
the gradient flow~\cite{Suzuki:2013gza}.

The gradient flow for the YM field $A_{\mu}$
in Euclidean space is a continuous transformation
of the gauge field according to the flow equation~\cite{Luscher:2010iy}\footnote{
  For the gradient flow for a fermion field,
  see Refs.~\cite{Luscher:2013cpa,Makino:2014taa}.},
\begin{align}
  \frac{dA_\mu(t,x)}{dt}
  = -g_0^2 \frac{\delta S_{\mathrm{YM}}(t)}{\delta A_\mu (t,x)},
  \label{eq:flow}
\end{align} 
where the flow time $t$ is a parameter 
controlling the magnitude of the transformation.
The YM action $S_{\mathrm{YM}}(t)$ is composed of $A_{\mu}(t,x)$,
whose initial condition at $t=0$ is
the ordinary gauge field $A_\mu(x)$ in the four dimensional
Euclidean space.
The gradient flow for positive $t$ smooths  the gauge field
with the radius $ \sqrt{2t}$.

Using the flowed field, 
the renormalized EMT operator in Euclidean space
is defined as~\cite{Suzuki:2013gza}
\begin{align}
  T_{\mu\nu}^{\rm E}(x) =& \lim_{t\rightarrow0} T_{\mu\nu}^{\rm E}(t,x),
  \label{eq:EMT}
  \\
  T_{\mu\nu}^{\rm E}(t,x) =& c_1(t) U_{\mu\nu}(t,x)
  \nonumber \\
  &+ c_2(t) \delta_{\mu\nu}  [ E(t,x)- \langle E(t,x) \rangle_0 ],
  \label{eq:EMTt}
\end{align}
where 
\begin{align}
  E(t,x) &= G_{\mu\nu}^a(t,x)G^a_{\mu\nu}(t,x) ,
  \label{eq:E}
  \\
  U_{\mu\nu}(t,x) &=
  G_{\mu\rho}^a(t,x)G_{\nu\rho}^a(t,x)-\frac14 \delta_{\mu\nu}E(t,x),
  \label{eq:U}
\end{align}
with the field strength $G_{\mu\nu}^a(t,x)$
composed of the flowed gauge field $A_\mu (t,x)$.
The vacuum expectation value $\langle T_{\mu\nu}^{\rm E}(t,x) \rangle_0$ is
normalized to be zero by the subtraction of $\langle E(t,x) \rangle_0$.
We use the perturbative coefficients $c_1(t)$ and $c_2(t)$ at two- and
three-loop orders~\cite{Suzuki:2013gza,Harlander:2018zpi,Iritani:2018idk},
respectively, in the following analysis~\cite{Iritani:2018idk}.
The EMT operator Eq.~(\ref{eq:EMT}) has been applied to the 
analysis of various observables in YM theories and
QCD with dynamical fermions~\cite{Asakawa:2013laa,Kamata:2016any,
Taniguchi:2016ofw,Kitazawa:2016dsl,Kitazawa:2017qab,Yanagihara:2018qqg,Hirakida:2018uoy}.
In particular, it has been shown that thermodynamics
in SU(3) YM theory is obtained accurately from the expectation values of
Eq.~(\ref{eq:EMT})~\cite{Kitazawa:2016dsl,Iritani:2018idk}.

In practical numerical simulations we measure
$\langle T_{\mu\nu}^{\rm E}(t,x)\rangle$ at nonzero $t$ and
lattice spacing $a$.
The flow time $t$ should be small enough to justify the use of the
perturbative coefficients for $c_1(t)$ and $c_2(t)$
as well as to suppress the oversmearing effect
which occurs when the operator is smeared larger than the temporal length~\cite{Kitazawa:2016dsl}.
In this range of $t$, the small flow time expansion~\cite{Luscher:2011bx}
implies that 
\begin{align}
  \langle T_{\mu\nu}(t,x)\rangle = \langle T_{\mu\nu}(x)\rangle + t c_{\mu\nu},
  \label{eq:tdep}
\end{align}
where $c_{\mu\nu}$ is a contribution from dimension six operators,
and contributions from yet higher dimensional operators are neglected.
As the lattice discretization effect on Eq.~(\ref{eq:tdep})
for $t>0$ is given by the powers of $a^2/t$~\cite{Fodor:2014cpa}
and diverges in the $t\to0$ limit,
the flow time must also satisfy $a\lesssim \sqrt{2t}$ to suppress
the discretization error.

\section{Numerical Setup}
\label{sec:setup}

\begin{table}
\begin{tabular}{r||l|r|r|l|r}
\hline \hline
$T/T_c$ & $\beta$ & $N_z$ & $N_\tau$ & $N_x$      & $N_{\rm vac}$ \\
\hline
1.12 & 6.418 &  72 & 12 &  12, 14, 16, 18     &  64 \\
     & 6.631 &  96 & 16 &  16, 18, 20, 22, 24 &  96 \\
\hline
1.40 & 6.582 &  72 & 12 &  12, 14, 16, 18     &  64 \\
     & 6.800 &  96 & 16 &  16, 18, 20, 22, 24 & 128 \\
\hline
1.68 & 6.719 &  72 & 12 &  12, 14, 16, 18, 24 &  64 \\
     & 6.719 &  96 & 12 &  14, 18             &  64 \\
     & 6.941 &  96 & 16 &  16, 18, 20, 22, 24 &  96 \\
\hline
2.10 & 6.891 &  72 & 12 &  12, 14, 16, 18, 24 &  72 \\
     & 7.117 &  96 & 16 &  16, 18, 20, 22, 24 & 128 \\
\hline
2.69 & 7.086 &  72 & 12 &  12, 14, 16, 18     & -\\
\hline
$\simeq8.1$
     & 8.0   &  72 & 12 &  12, 14, 16, 18     & -\\
\hline
$\simeq25$
     & 9.0   &  72 & 12 &  12, 14, 16, 18     & -\\
\hline \hline
\end{tabular}
\caption{
  Simulation parameters $\beta=6/g_0^2$ and lattice volume
  $N_x\times N_z^2 \times N_\tau$ for each temperature $T$.
  The vacuum subtraction is performed on lattices with $N_{\rm vac}^4$.}
\label{table:param}
\end{table}

We have performed numerical simulations of SU(3) YM theory
on four-dimensional Euclidean lattices with the PBC for all directions.
The simulations are performed with the standard Wilson gauge action
for an isotropic lattice~\cite{Rothe:1992nt}\footnote{Note that here isotropy refers to the equal spacing between all lattice points, as was done in this work.}
for several values of $\beta=6/g_0^2$
and the lattice volume $N_x\times N_z^2 \times N_\tau$ summarized in 
Table~\ref{table:param}.
The lattice spacing $a$ and temperature $T$ are determined according
to the relation between $\beta$ and $a$ in Ref.~\cite{Kitazawa:2016dsl}.
The lattice size along $y$ and $z$ directions is fixed to $N_z/N_\tau=6$,
except for the $N_x\times 96^2\times12$ lattices at $T/T_c=1.68$
used for the analysis of the dependence on $N_z/N_\tau$ in Sec.~\ref{sec:Nz}.
In the conventional analysis of the isotropic
thermodynamics on lattices with
$N_s^3\times N_\tau$, it is practically known that the finite-size effect
is well suppressed at the aspect ratio $N_s/N_\tau=4$~\cite{Boyd:1996bx}.
The ratio $N_z/N_\tau=6$ in our simulations is 
larger than this value\footnote{In our simulation $N_x/N_\tau\sim1<4$ because we are explicitly interested in numerically determining the finite-size corrections.}.
For the vacuum subtraction,
we use the data obtained on $N_{\rm vac}^4$ lattices.
Except for the simulation at $\beta=6.891$, 
we use the data used in Ref.~\cite{Kitazawa:2016dsl}.

As our code cannot deal with odd $N_x$,
we have performed the analyses for even number of $N_x$
shown in Table~\ref{table:param}.
Under this constraint,
it is difficult to perform the simulations
at the same lattice volume $L_x\times L_z^2$ and $T$
with different $a$ 
in general.
Therefore, 
in the present study we do not take the continuum extrapolation.
Instead, we perform numerical analyses with two different lattice
spacings at $N_\tau=12$ and $16$ for $1.12\le T/T_c \le 2.1$ to investigate
the lattice discretization effect, which will be discussed
in Sec.~\ref{sec:t->0}.
We restrict ourselves to $T>T_c$ in the present study,
as the results for $T<T_c$ currently have statistical errors
too large to draw meaningful conclusions.

We perform $2,100-4,000$ measurements for each set of parameters
at nonzero $T$.
Each measurement is separated by $100$ Sweeps, where one Sweep is
composed of one pseudo heat bath and five over relaxation
updates~\cite{Kitazawa:2016dsl}.
The number of measurements for the vacuum is $560-1,020$.
All statistical errors are estimated by the jackknife method with
binsize $20$, at which the binsize dependence of the 
statistical error is not observed.

Other procedures and the implementation of the simulation are
the same as those in Ref.~\cite{Kitazawa:2016dsl}.
We use the Wilson gauge action for $S_{\rm YM}(t)$ in the flow equation
Eq.~(\ref{eq:flow}).
For the operator $U_{\mu\nu}(t,x)$ in Eq.~(\ref{eq:U}),
we use $G^a_{\mu\nu}(t,x)$ written in terms of the clover-leaf
representation~\cite{Rothe:1992nt}.
For $E(t,x)$ in~Eq.~(\ref{eq:E}),
we use the tree-level improved
representation~\cite{Fodor:2014cpa,Kitazawa:2016dsl,Kamata:2016any},
 \begin{align}
E(t,x)_{\rm imp} = \frac34 E(t,x)_{\rm clover} + \frac14 E(t,x)_{\rm plaq},
\label{eq:E_imp}
\end{align}
where $E(t,x)_{\rm clover}$ is constructed from the clover-leaf 
representation of $G^a_{\mu\nu}(t,x)$ and
$E(t,x)_{\rm plaq} $ is defined from the plaquette~\cite{Luscher:2010iy}. 
We use the iterative formula for four-loop running coupling
\cite{Tanabashi:2018oca} and the value of $\Lambda_{\overline{\mathrm{MS}}}$
determined in Ref.~\cite{Kitazawa:2016dsl} 
for the perturbative coefficients $c_1(t)$ and $c_2(t)$.
This combination of the running coupling and the perturbative coefficients
at different orders is known to give a good description of thermodynamics~\cite{Iritani:2018idk}.
We estimate the systematic error from an uncertainty of
$\Lambda_{\overline{\mathrm{MS}}}$ by varying the value by $\pm3\%$
in the following unless otherwise stated.

\section{Numerical results}
\label{sec:result}

\subsection{\texorpdfstring{$t\to0$}{t->0} extrapolation}
\label{sec:t->0}

\begin{figure}
  \includegraphics[width=0.49\textwidth,clip]{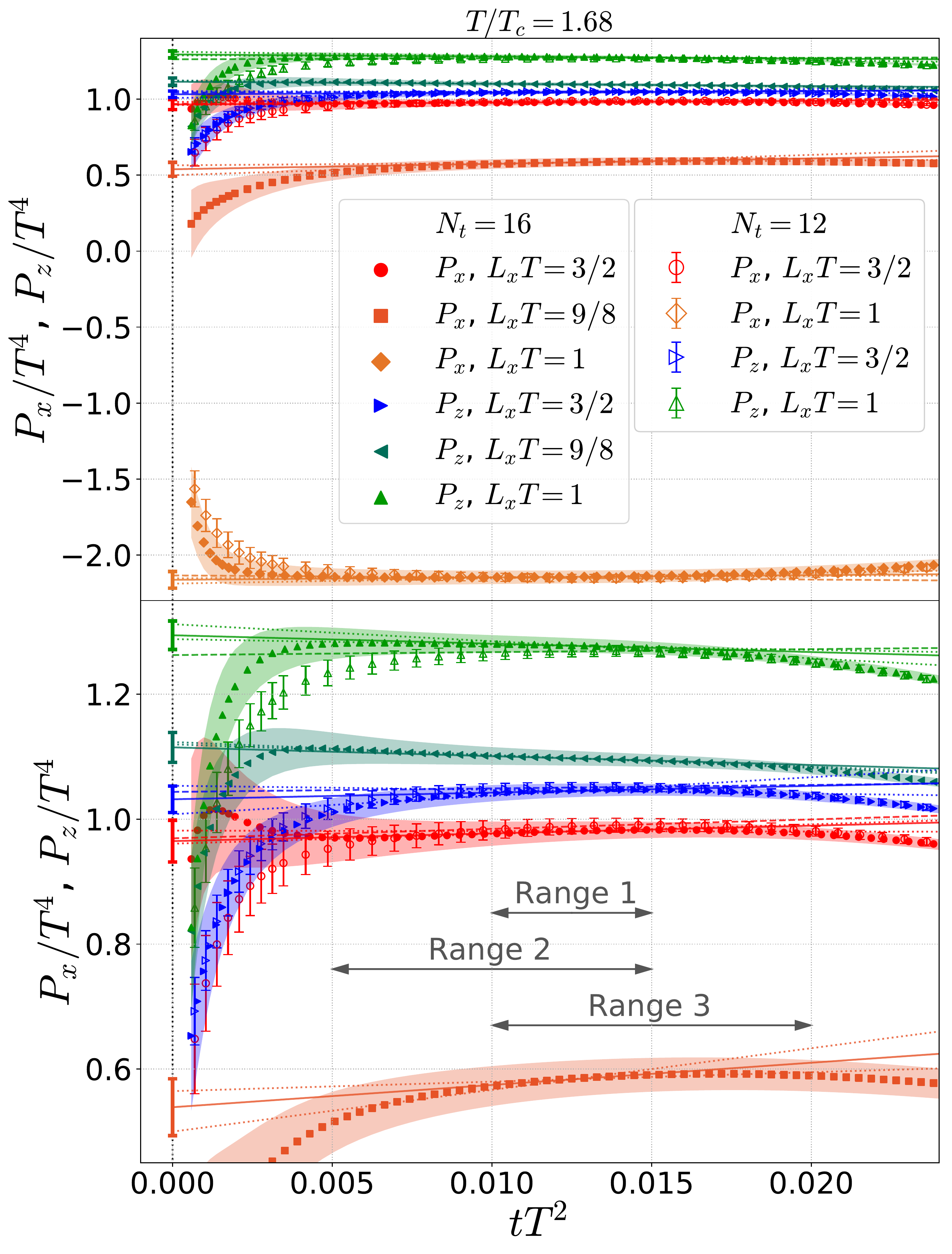}
\caption{
  Flow time $t$ dependences of $P_x(t)/T^4$ and $P_z(t)/T^4$
  for $T/T_c=1.68$ with $N_\tau=16,~12$ and $L_xT=1,~9/8,~3/2$.
  Statistical errors for $N_\tau=16$ are shown by the shaded area,
  while those for $N_\tau=12$ are shown by error bars.
  Lower panel is an expansion of upper panel.
  Solid (dashed) lines show the $t\to0$ extrapolation obtained
  with the data for $N_\tau=16$ ($12$) in Range-1.
  Dotted lines show the extrapolations with Range-2 and Range-3
  with $N_\tau=16$.
  Extrapolated results of $P_x/T^4$ and $P_z/T^4$ to $t\to0$ with Range-1
  at $N_\tau=16$ and their statistical errors 
  are shown on the $tT^2=0$ axis.}
\label{fig:166}
\end{figure}

We first focus on the result for $T=1.68T_c$ and discuss the
$t$ and $a$ dependences of the numerical results.
In Fig.~\ref{fig:166}, we show the $t$ dependence of
$P_x(t)=\langle T_{11}(t,x) \rangle$ and
$P_z(t)=\langle T_{22}(t,x) \rangle =\langle T_{33}(t,x) \rangle$
at $T/T_c=1.68$ and $L_x T=N_x/N_\tau=1$, $9/8$, and $3/2$.
The lower panel is a magnified plot of the upper panel for the
range $0.45\le P_x(t)/T^4 , P_z(t)/T^4 \le 1.35$.
For $L_x T=1$ and $3/2$, we show results for two lattice spacings,
$N_\tau=16$ (filled symbols) and $12$ (open symbols).
The statistical errors are shown by the shaded area (error bars)
for $N_\tau=16$ ($N_\tau=12$).
From Fig.~\ref{fig:166}, one finds that $P_x(t)$ and $P_z(t)$ behave
almost linearly as functions of $t$ in the range
$0.005\lesssim tT^2\lesssim 0.02$~\cite{Kitazawa:2016dsl,Iritani:2018idk}\footnote{
  As $T$ and $a$ are related with each other as $a=(N_\tau T)^{-1}$,
  the lower boundary of this condition corresponds to
  $0.005\lesssim t(aN_\tau)^{-2}$.
  For $N_\tau=12$, we thus have $0.72\lesssim t/a^2$, which is consistent
  with the argument below Eq.~(\ref{eq:tdep}).}.
The deviations from this behavior at small and large $t$ come from
lattice discretization and oversmearing effects,
respectively~\cite{Kitazawa:2016dsl}.

The expectation value of the EMT is obtained by taking the $t\to0$ limit
of these results.
In Refs.~\cite{Kitazawa:2016dsl,Iritani:2018idk},
the $t\to0$ limit is taken after
the continuum extrapolation for each value of $t$.
From the data sets in the present study, however, the continuum
extrapolation cannot be taken because we do not have the results 
with different lattice spacings with the same volume $L_x\times L_z^2$
except for $L_x T=1$ and $1.5$.
We thus take the $t\to0$ limit for each $N_\tau$
assuming a linear $t$ dependence Eq.~(\ref{eq:tdep}).
For the fitting range of the extrapolation, we employ three
ranges~\cite{Kitazawa:2016dsl,Iritani:2018idk}:
\begin{description}
 \item[Range-1] $ 0.01 \le tT^2 \le 0.015$, 
 \item[Range-2] $ 0.005 \le tT^2 \le 0.015$, 
 \item[Range-3] $ 0.01 \le tT^2 \le 0.02$,
\end{description}
which are shown in the lower panel of Fig.~\ref{fig:166} by the arrows.
The $t\to0$ extrapolation for $N_\tau=16$ with Range-1
is shown by the solid line in Fig.~\ref{fig:166}, while the 
extrapolated values of $P_x/T^4$ and $P_z/T^4$ are plotted on 
the $tT^2=0$ axis with the statistical error.
The fitting results for $N_\tau=16$ with Range-2 and Range-3
are shown by the dotted lines.
We use the result with Range-1 as a central value, while those with
Range-2 and Range-3 are used to estimate the systematic error
associated with the fitting range.
As Fig.~\ref{fig:166} shows, this systematic error is 
at most comparable with the statistical one in spite of
the large variation of the fit range~\cite{Kitazawa:2016dsl}.
In Fig.~\ref{fig:166}, the $t\to0$ extrapolation for $N_\tau=12$
with Range-1 is also shown by the dashed lines for $L_xT=1$ and $3/2$.

Comments on the $t\to0$ extrapolation are in order.
First, unlike the analysis in Refs.~\cite{Kitazawa:2016dsl,Iritani:2018idk},
the results in the present study are not the continuum extrapolated one.
However, the numerical results in this analysis are expected to be
close to these after the continuum extrapolation
because of the following reasons.
First, when the lattice spacing becomes finer, 
our analysis converges to the continuum extrapolated analysis
in Refs.~\cite{Kitazawa:2016dsl,Iritani:2018idk}, as 
the difference is proportional to $a^2$ for sufficiently small $a$.
Second, the discretization effect is expected
to be well suppressed already at $N_\tau=12$.
In fact, Fig.~\ref{fig:166} shows that the values of $P_x(t)$ and $P_z(t)$
for $N_\tau=16$ and $12$ at $L_xT=1$ and $3/2$ agree with each other
within statistics for $0.005\le tT^2$.
As a result, the $t\to0$ extrapolated values $P_x$ and $P_z$ also agree
within statistics.
Furthermore, we performed the analysis of the data at $N_\tau=12$ and $16$
in Ref.~\cite{Kitazawa:2016dsl} by the method in the present study,
and compared them with the continuum extrapolated results
in Ref.~\cite{Iritani:2018idk}.
From this analysis we have checked that the results
agree with each other within $2\sigma$ for $1.12\le T/T_c\le2.1$.
Therefore, given the uncertainty in the $t\rightarrow0$ extrapolation, the lattice spacing is expected to be sufficiently small for suppressing the discretization effects of $\langle T_{\mu\nu}^{\rm E} \rangle$
already at $N_\tau=12$ and $16$.

\subsection{\texorpdfstring{$N_z/N_\tau$}{N\_z/N\_tau} dependence}
\label{sec:Nz}

\begin{figure}
  \includegraphics[width=0.49\textwidth,clip]{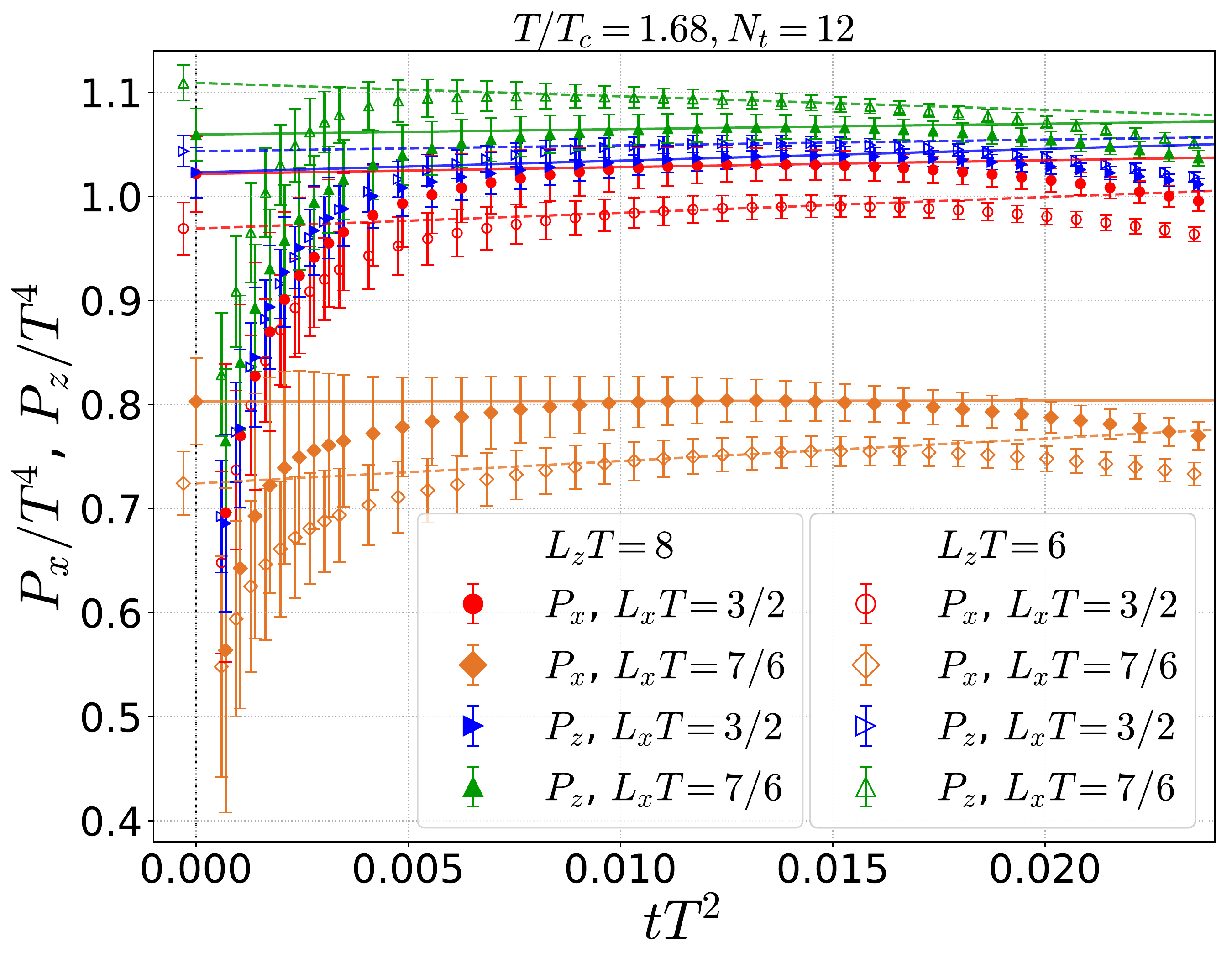}
\caption{
  Flow time $t$ dependences of $P_x(t)/T^4$ and $P_z(t)/T^4$
  for $T/T_c=1.68$ and $N_\tau=12$ with different values of
  $L_zT=N_z/N_\tau$.
  Data points at $L_zT=6$ are shifted toward left slightly.
  Solid (dashed) lines show the $t\to0$ extrapolation for 
  $L_zT=8$ (6) with Range-1.
  Extrapolated values of $P_x/T^4$ and $P_z/T^4$ are shown
  around $tT^2=0$ axis with their statistical error.}
\label{fig:Nz}
\end{figure}

We wish to study the finite-size corrections in lattice simulations of thermodynamic properties when only one direction is of finite size, in this case the $x$ direction. Since our calculations are performed on the lattice, the $y$ and $z$ directions are necessarily finite.  We would therefore like to see that our results are insensitive to this finite-size in the $y$ and $z$ directions.  As noted previously, finite-size effects are small in isotropic lattices with $N_s/N_\tau=4$~\cite{Boyd:1996bx}.  All our results were found using $L_zT=N_z/N_\tau=6$, so we expect any finite-size effects in the $y$ and $z$ directions to be well suppressed.  To test this hypothesis, we perform a numerical analysis with $L_zT=8$ at
$N_x=14$ and $18$ for $T/T_c=1.68$ and $N_\tau=12$ and compare to our usual $L_zT=6$ results.  
In Fig.~\ref{fig:Nz} we compare the $t$ dependences of
$P_x(t)$ and $P_z(t)$. 
(The number of measurements for $L_zT=8$ is $1,000$.)
The $t\to0$ extrapolation with Range-1 is shown by the
solid (dashed) lines for $L_zT=8$ ($L_zT=6$), with the extrapolated values of $P_x$ and $P_z$ shown around $tT^2=0$.
As can be seen in the figure, the values of $P_x$ and $P_z$ thus obtained
for $L_zT=8$ and $6$ agree within $\lesssim1\sigma$ of their statistical errors.
These results suggest that the boundary effect along the $y$ and $z$
directions in our lattice simulations is well suppressed,
while the data at nonzero $tT^2$ in Fig.~\ref{fig:Nz} might suggest
the existence of the $L_zT$ dependence at $L_zT=6$ which should be
studied by the future numerical analysis with much higher statistics.

\subsection{Pressure anisotropy}
\label{sec:pressure}

\begin{figure}
  \includegraphics[width=0.49\textwidth,clip]{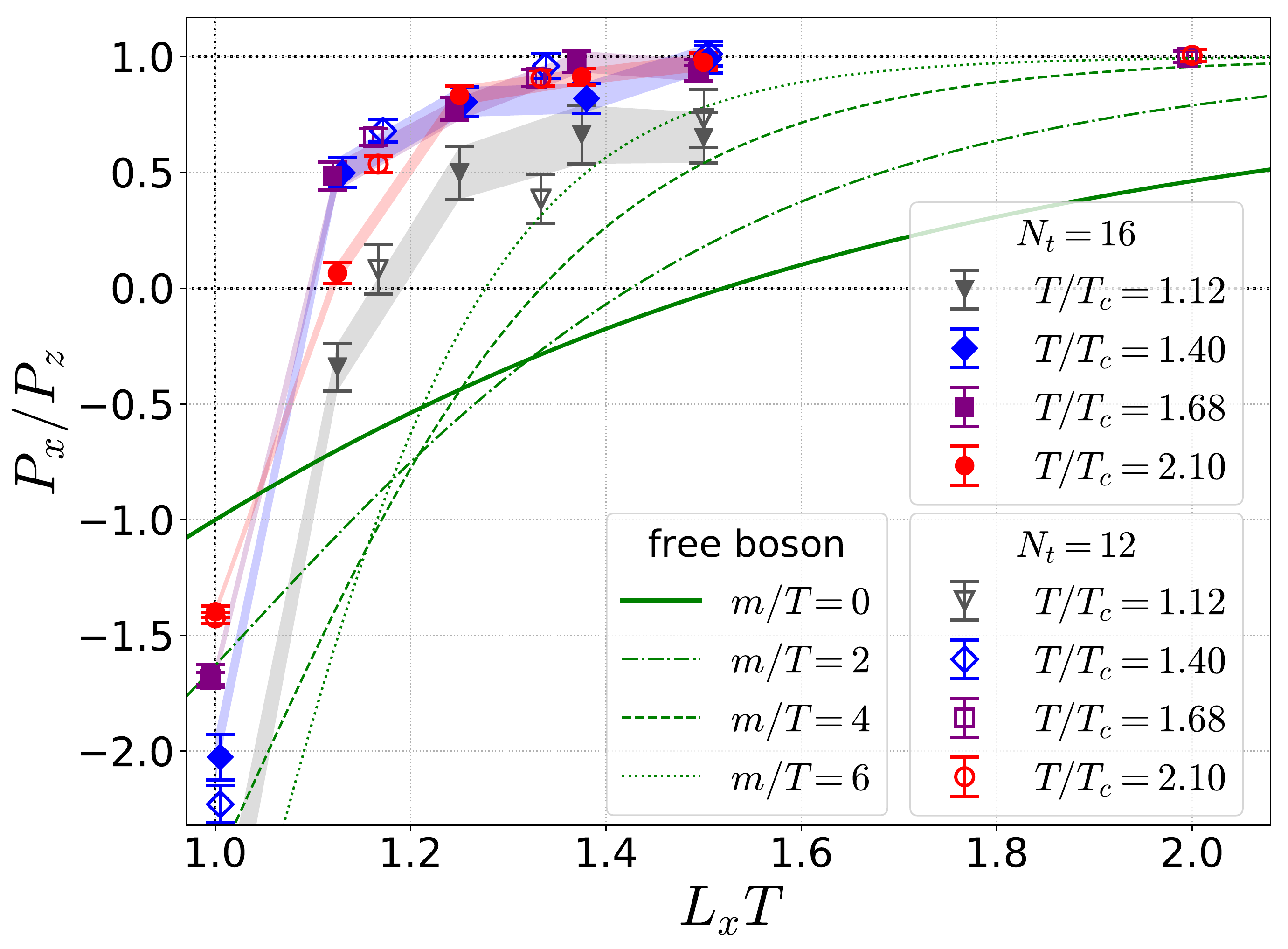}
\caption{
  Ratio $P_x/P_z$ as a function of $L_xT$ for various values of $T/T_c$.
  Error bars include statistical error and systematic ones from
  (1) the choice of the fit range and
  (2) $\pm3\%$ uncertainty of $\Lambda_{\overline{\mathrm{MS}}}$; see the text.
  The behavior of $P_x/P_z$ in the free scalar theory is also shown
  by the lines for several values of mass temperature ratio $m/T$.
  Shaded bands connect error bars at $N_\tau=16$.
  The data points at $T/T_c=1.40$ (1.68) are shifted toward right (left)
  slightly.}
\label{fig:ratio}
\end{figure}

Now, let us first focus on the ratio $P_x/P_z$.
In Fig.~\ref{fig:ratio}, we show the $t\to0$ extrapolated results
of $P_x/P_z$ as a function of $L_xT$
at four temperatures, $T/T_c=1.12$, $1.40$, $1.68$, and $2.10$.
The results for $N_\tau=16$ and $12$ are shown by the filled and open symbols,
respectively.
Error bars include systematic error from the choice of the fitting range
and the uncertainty of $\Lambda_{\overline{\mathrm{MS}}}$ estimated from 
$\pm3\%$ variation, as well as the statistical one.
The comparison of the results for $N_\tau=16$ and $12$ shows that
a significant lattice spacing dependence is not observed,
as anticipated from the discussion in Sec.~\ref{sec:t->0}.

In Fig.~\ref{fig:ratio}, we also show the ratio $P_x/P_z$ obtained in
the free scalar theory with mass $m$ for several values of $m/T$.
The result for $m=0$ is taken from Ref.~\cite{Mogliacci:2018oea},
while the procedure to obtain the results at $m\ne0$ will be
reported in a future publication~\cite{Mogliacci:prep}.

As discussed in Sec.~\ref{sec:aniso}, $P_x/P_z$ approaches unity
in the $L_xT\to\infty$ limit.
In the free massless theory, 
a clear deviation of $P_x/P_z$ from this limiting value is already
observed at $L_xT=2$, and the ratio crosses zero at $L_xT\simeq1.5$.
At $L_xT=1$, the ratio is $P_x/P_z=-1$, as suggested from
Eq.~(\ref{eq:px=e}) and the fact that $\Delta=0$ in this theory.

The results of SU(3) YM theory shown in Fig.~\ref{fig:ratio}
behave quite differently from the massless free theory.
In SU(3) YM theory, $P_x/P_z=1$ within statistics at $L_xT=1.5$
for $1.4\le T/T_c \le2.1$.
Even at $L_xT=1.333$ and $1.375$, deviation from $P_x/P_z=1$ is
comparable with the error for these temperatures.
By decreasing $L_xT$ further, the ratio suddenly becomes smaller 
and arrives at $P_x/P_z<-1$ at $L_xT=1$.
It is interesting to note that almost the same $L_xT$ dependence
is observed for $1.4\le T/T_c \le2.1$, while the result near $T_c$
at $T/T_c=1.12$ shows a deviation from this trend.
From these results, it is concluded that the SU(3) YM theory
at $1.4\le T/T_c \le2.1$ is remarkably insensitive to the PBC with length $L_x$ compared with the massless free theory.
At $T/T_c=1.12$, 
the SU(3) YM theory is however clearly more sensitive to the PBC.
This may be important for future phenomenological applications.

In the free scalar theory, $P_x/P_z$ approaches unity as $m/T$ becomes
larger for large $L_xT$ as shown in Fig.~\ref{fig:ratio}.
Therefore, the lattice results might be partially understood as 
the effect of nonzero $m$. However, even at $m/T=6$ the behavior is
still inconsistent with the lattice result.
Also, $P_x/P_z$ at $L_xT=1$ becomes smaller as $m/T$ becomes larger,
which is inconsistent with the lattice result.

\begin{figure*}
  \includegraphics[width=0.49\textwidth,clip]{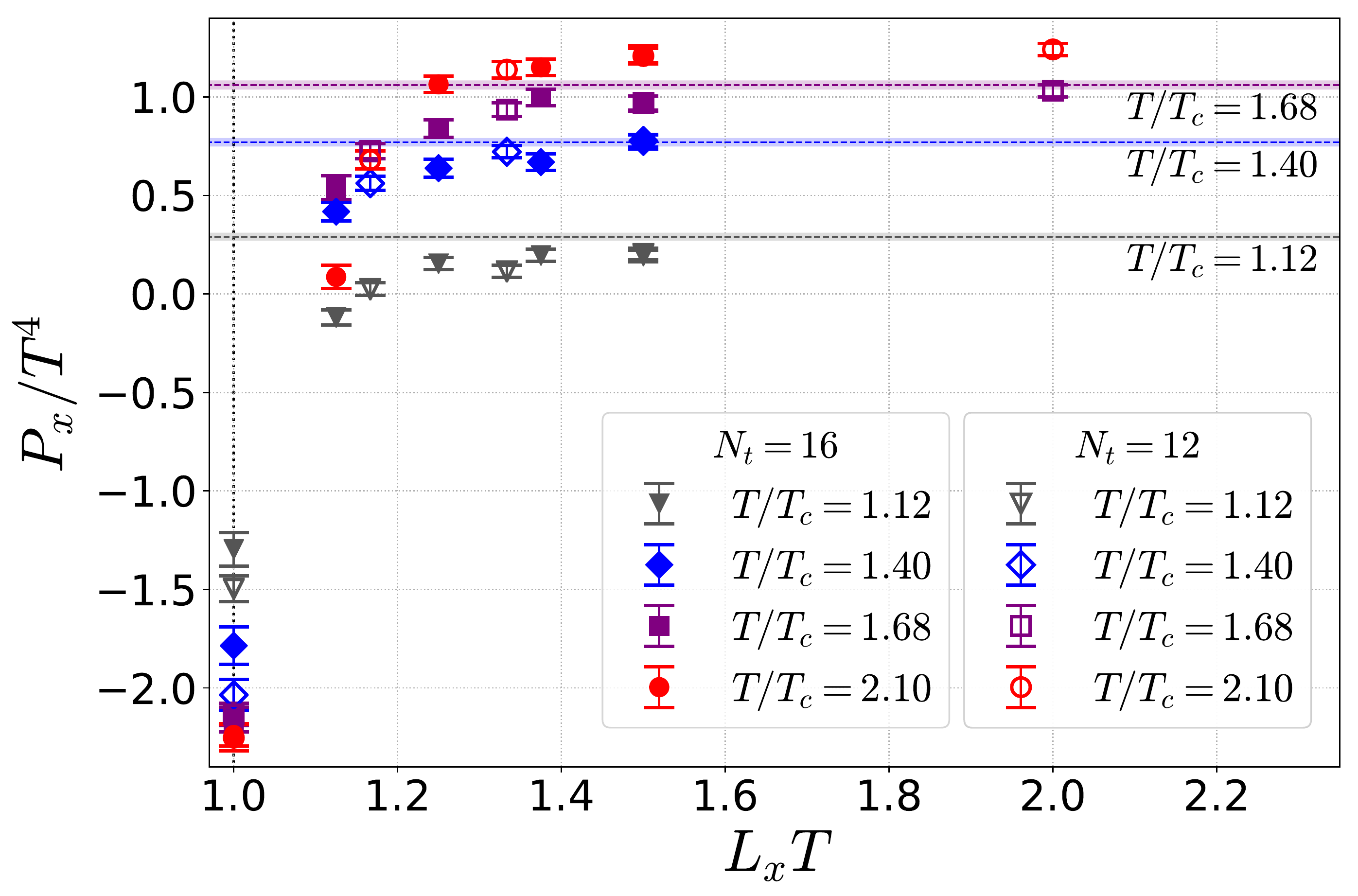}
  \includegraphics[width=0.49\textwidth,clip]{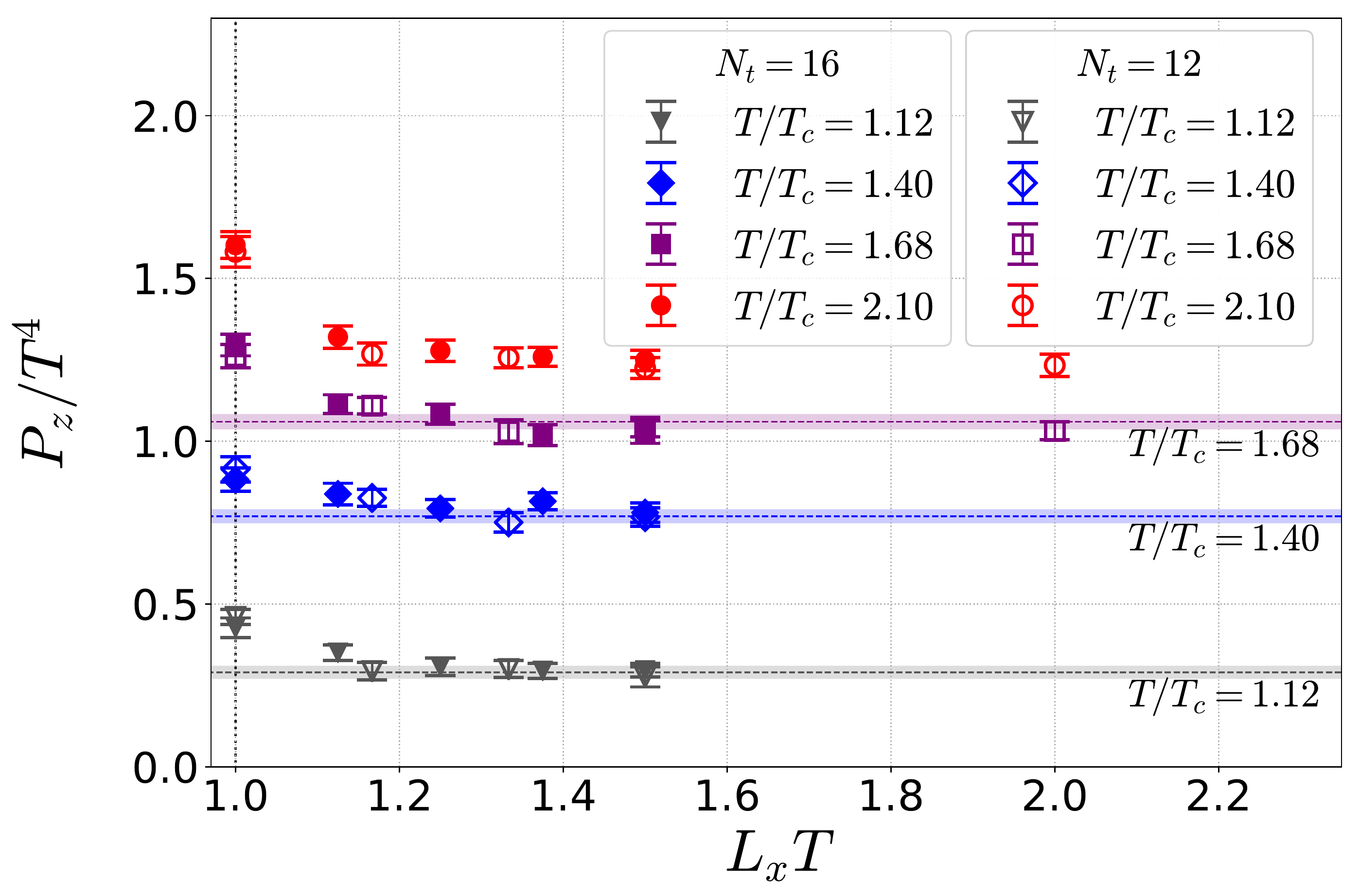}
  \includegraphics[width=0.49\textwidth,clip]{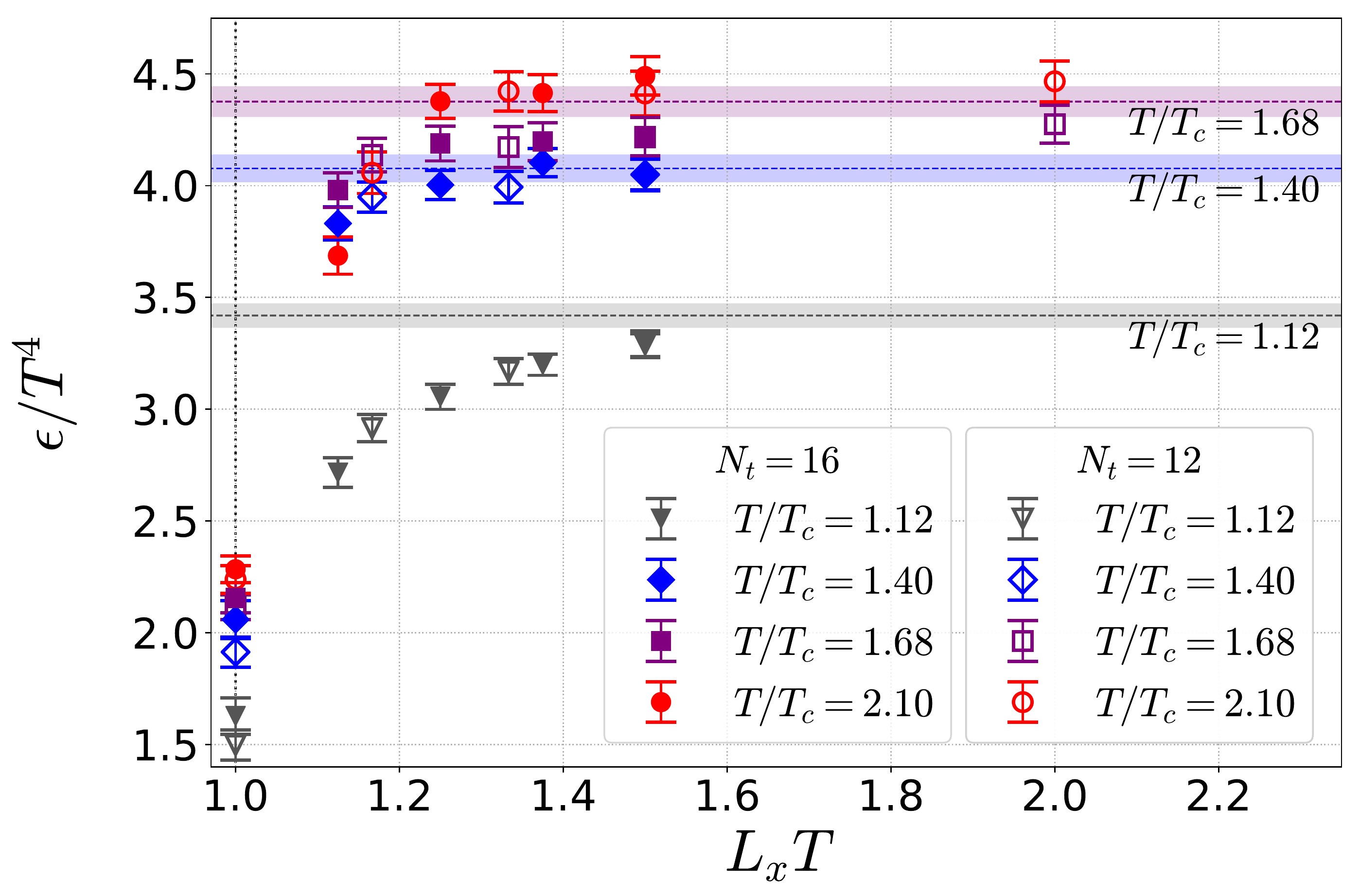}
  \includegraphics[width=0.49\textwidth,clip]{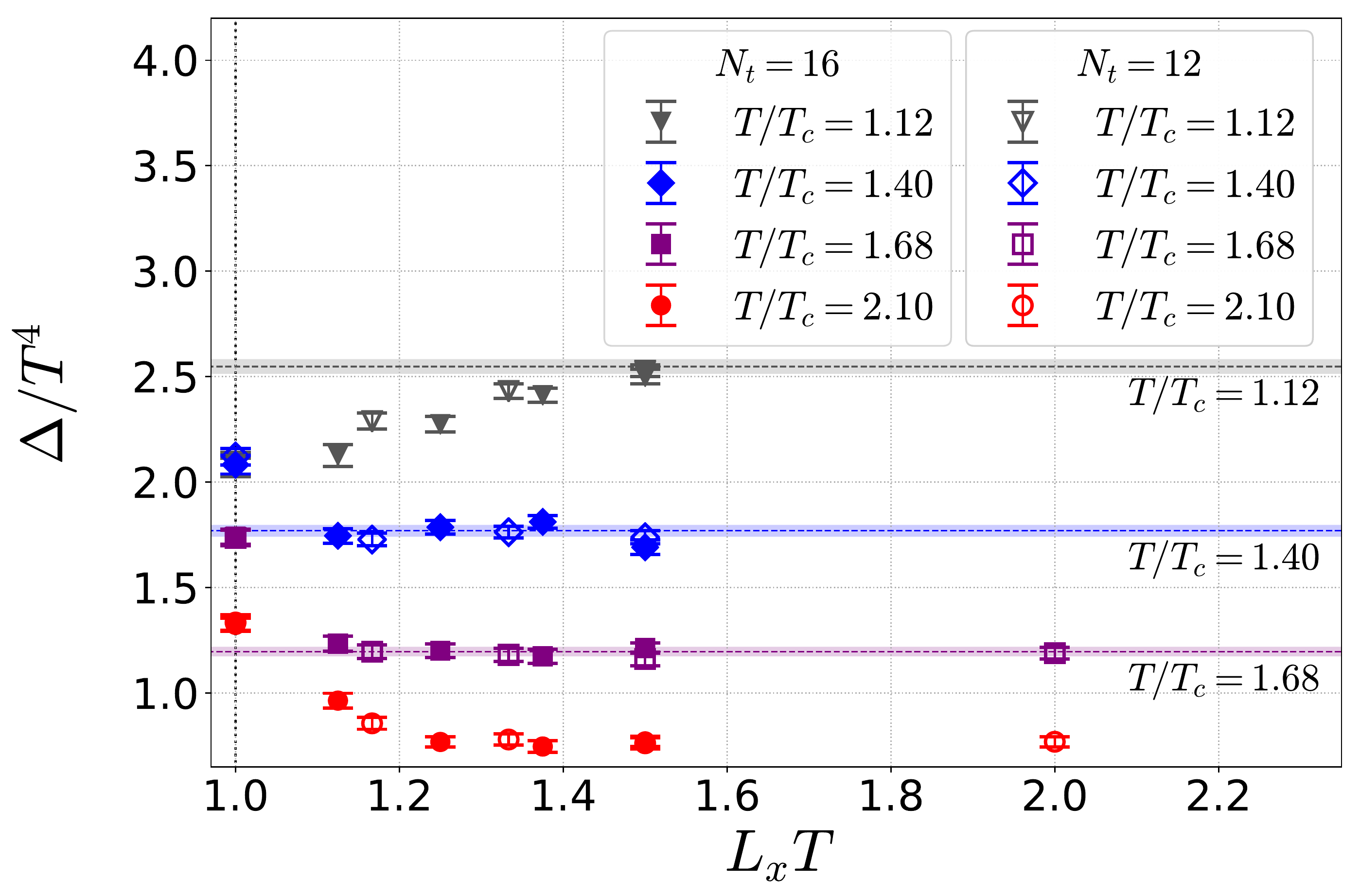}
\caption{
  Dependences of $P_x/T^4$ (upper-left),
  $P_z/T^4$ (upper-right), $\varepsilon/T^4$ (lower-left), and
  $\Delta/T^4$ (lower-right)
  on $L_xT$ for several values of $T/T_c$.
  Dashed vertical lines in the upper (middle and lower) panel(s) show
  the value of $\varepsilon/T^4$ ($P/T^4$) in the isotropic case
  obtained in Ref.~\cite{Iritani:2018idk}, with the error shown 
  by the shaded region.}
\label{fig:epz}
\end{figure*}

Shown in Fig.~\ref{fig:epz} are 
the behavior the longitudinal and transverse pressures $P_x$ and $P_z$,
the energy density $\varepsilon$, and $\Delta$
as functions of $L_x T$.
For guides of these results, we also show the continuum
extrapolated values of $P/T^4$, $\varepsilon/T^4$, and $(\varepsilon-3P)/T^4$
in the isotropic case obtained in Ref.~\cite{Iritani:2018idk}
by the horizontal dashed lines for $T/T_c=1.12,~1.40,~1.68$,
with the errors shown by the shaded region.
From Fig.~\ref{fig:epz}, one finds that these quantities 
are insensitive to the existence of the boundary for
$L_xT\gtrsim1.3$ for $T/T_c=1.40,~1.68,~2.10$.

\subsection{High temperature}
\label{sec:highT}

\begin{figure}
  \includegraphics[width=0.49\textwidth,clip]{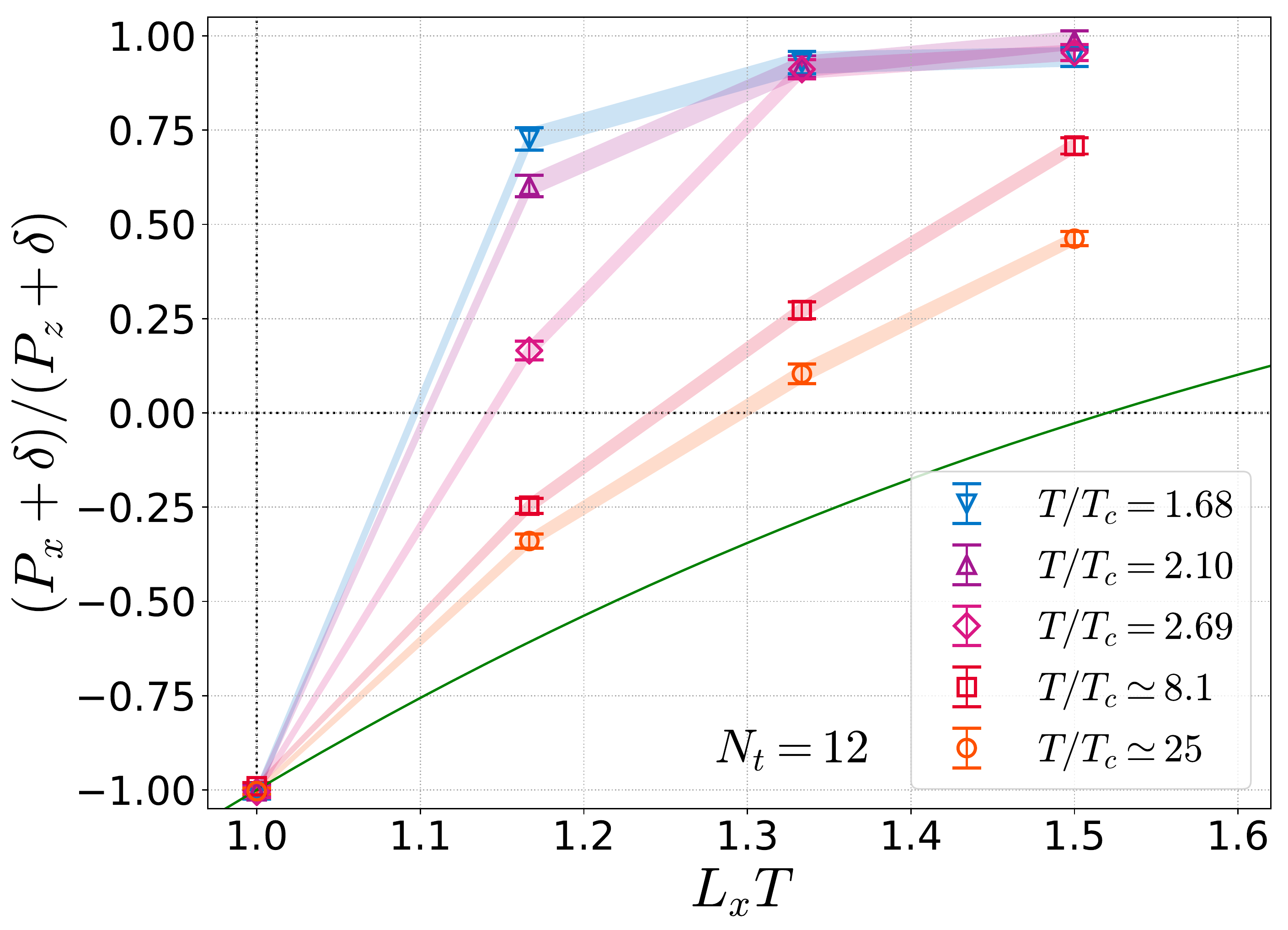}
  \includegraphics[width=0.49\textwidth,clip]{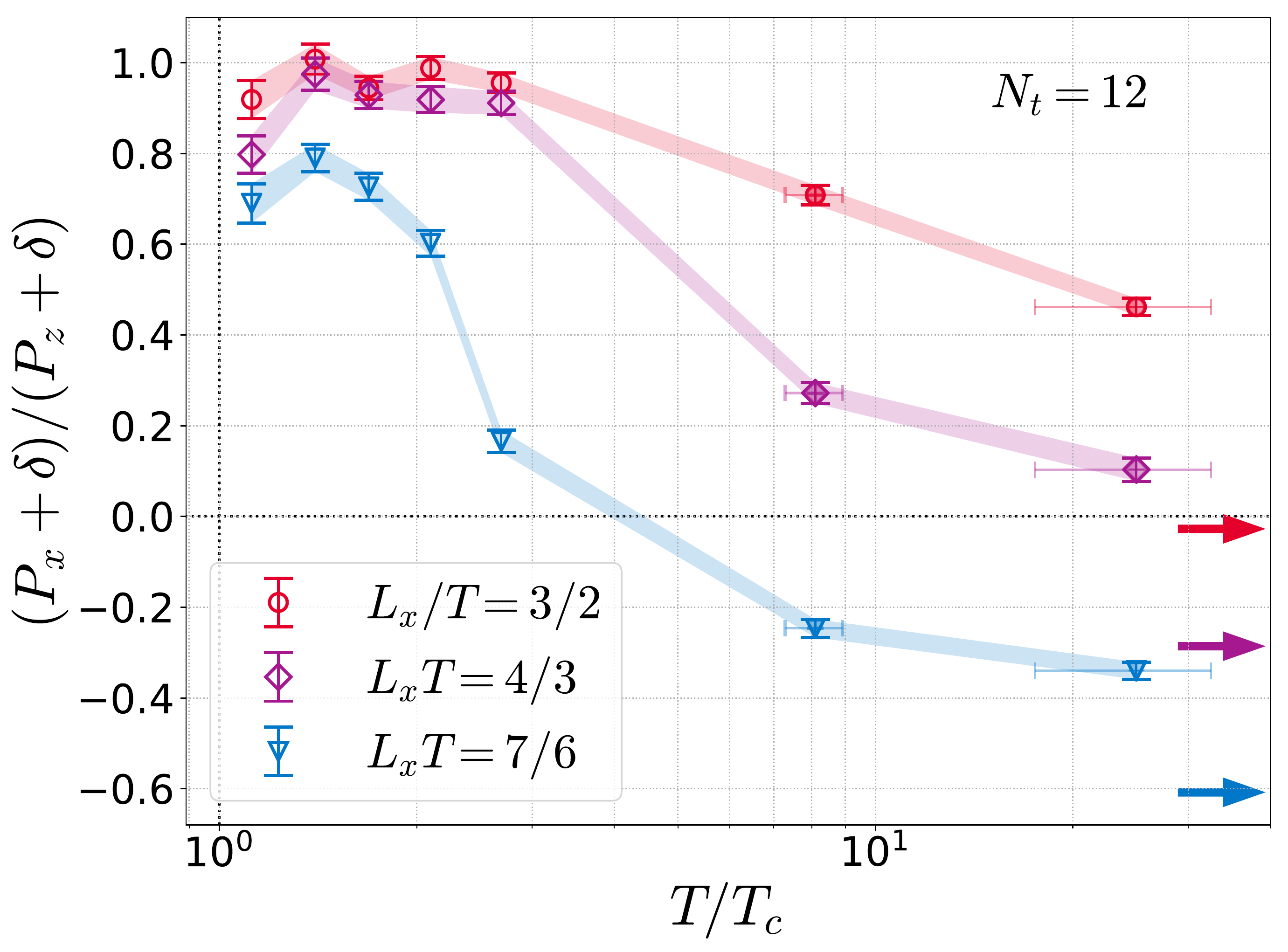}
\caption{
  Ratio $(P_x+\delta)/(P_z+\delta)$ for various values of $T$ and
  $L_xT$. Upper (Lower) panel shows the ratio as a function of
  $L_xT$ ($T/T_c$).
  The solid line in the upper panel shows the ratio in the
  massless free scalar theory.
  The arrows at the right in the lower panel shows the ratio in the
  massless free scalar theory for each $L_xT$.}
\label{fig:highT}
\end{figure}

At asymptotically high temperature, the SU(3) YM theory 
approaches a free gas composed of massless gluons.
In this limit, the $L_xT$ dependence of $P_x/P_z$ should approach 
the massless free scalar theory.
It is an interesting question how the results in Fig.~\ref{fig:ratio}
approach this asymptotic behavior.
The extension of the numerical analysis to high $T$, however,
has two difficulties.
First, as the lattice spacing $a=(N_\tau T)^{-1}$ becomes smaller
the lattice size required for the vacuum subtraction becomes huge.
Second, the relation between $\beta=6/g_0^2$ and $a$ is not
available for such fine lattice spacings.

Here, to extend our analyses to high temperatures avoiding
these difficulties we focus on the ratio
\begin{align}
  \frac{P_x+\delta}{P_z+\delta},
  \label{eq:ratioD}
\end{align}
with $\delta=\Delta/4$.
This ratio does not depend on the second term in Eq.~(\ref{eq:EMTt})
proportional to $c_2(t)$.
One thus can obtain the ratio without the vacuum subtraction.
Furthermore, as $c_1(t)$ cancels between numerator and denominator
in Eq.~(\ref{eq:ratioD}), this ratio is obtained without using $c_1(t)$.
This means that the lattice spacing in physical
 units required for the determination of the running coupling in $c_1(t)$~\cite{Iritani:2018idk}
is not needed to obtain Eq.~(\ref{eq:ratioD}).

In Fig.~\ref{fig:highT}, we show the behavior of Eq.~(\ref{eq:ratioD})
as functions of $L_xT$ and $T/T_c$ in the upper and lower panels,
respectively.
The results at $T/T_c\simeq8.1$ and $25$ corresponds to those obtained
at $\beta=8.0$ and $9.0$, respectively; see Table~\ref{table:param}.
Temperatures are deduced from the relation between $\beta$ and
$a$ in Ref.~\cite{Kitazawa:2016dsl}, which is reliable
for $6.3\le\beta\le7.4$.
As $\beta=8.0$ and $9.0$ are outside of this range,
the values $T/T_c$ should be regarded just as a guide for the true
value of $T/T_c$.
To depict this uncertainty, in the lower panel we show $10\%$ and
$30\%$ error bars in $T/T_c$ for the data points at $\beta=8.0$ and $9.0$.

In the upper panel of Fig.~\ref{fig:highT}, we show
the ratio Eq.~(\ref{eq:ratioD}) in the massless free scalar theory
by the solid line, while in the lower panel the ratio for each
$L_xT$ is shown by arrows at right in the panel;
note that in the massless theory $\delta=0$.
The comparison of the lattice data with these results shows that 
the former approaches the asymptotic value as $T$ is increased,
but the difference is still large even at the highest temperature
$T/T_c\simeq25$.

\section{Discussion and Outlook}
\label{sec:outlook}

In the present study, we investigated the energy momentum tensor in 3+1 dimensional SU(3) YM theory at $T>T_c$ in anisotropic finite volume systems with the PBC.  We chose to make one direction small, $L_xT\sim1$, while keeping the other two spatial dimensions large, $L_{y,z}T\gg1$.
We found that, as shown in Fig.~\ref{fig:ratio},
a clear anisotropy in the stress tensor is observed
only for $L_xT\lesssim 1.3$ for $1.4\le T/T_c\le2.1$.
In free scalar theory with the same boundary condition,
a significant anisotropy manifests itself at much larger values of $L_xT$.
One therefore concludes that SU(3) YM theory near but above $T_c$ is
remarkably insensitive to the existence of the periodic boundary.
Even allowing the free scalar particles to have a mass $m=6T$ was insufficient to reproduce the insensitivity to the presence of the finite periodic boundary in SU(3) YM theory.

At the scales probed by these temperatures the running coupling is $g(2\pi T)\sim2$, and the leading order, infinite volume thermal field theory result for the Debye mass of the gluon is $m_D\sim gT$.  That the effective free quasiparticle mass required to mimic the results of the full SU(3) YM theory is so large indicates that 
1) finite-size corrections to the infinite and isotropic volume leading order thermal field theory result are large, for example the Debye gluon mass --- which by dimensional analysis is given by $m_D/g=f_T(L_xT) T +f_{L_x}(L_xT)/L_x$ --- might pick up large finite-size corrections, 
2) the interactions of the full theory cannot be easily approximated by a free quasiparticle theory, or 3) that there are important non-perturbative dynamics at these scales.

Investigating 1) is an important avenue for future analytic research, especially as the work here possibly suggests that the finite-size corrections to the effective gluon mass are large. 2) is quite likely given than other thermodynamic properties computed from the lattice at these temperature scales are only well approximated by resummed thermal field theory at three or four loops \cite{Andersen:2012wr,Mogliacci:2013mca,Haque:2014rua}. 3) must also contribute: Forty years ago, Linde demonstrated~\cite{Linde:1980ts} the possibility for an infrared cutoff of order ${\cal O}\left(g^2 T\right)$ to appear in the thermodynamics of a YM gas in an isotropic infinite volume. This effectively led to the findings of a non-perturbative coefficient in the pressure, when probed perturbatively~\cite{DiRenzo:2006nh}. More recently, the presence of the very same type of (Linde) problem was discovered in an anisotropic volume of SU(3) YM theory~\cite{Fraga:2016oul}, such as the one we use here. These works obviously raise the need for a better understanding of the possible presence of a non-perturbative scale such as $\sim g \sqrt{T/L_x}$ in the thermodynamics of anisotropic volumes of the SU(3) YM theory.
It is then an interesting future work to pursue the physical origin from the knowledge of the Casimir effect in various theories and settings~\cite{Meyer:2009kn,Karabali:2018ael,Chernodub:2018aix,Ishikawa:2018yey,Chernodub:2019nct}.

The remarkably large effective quasiparticle mass required to mimic the lattice results suggests a larger-than-expected effective Debye mass for gluons at temperatures on the order of $T_c$.  A larger Debye mass implies a stronger-than-expected screening of color charges in the thermal medium, which would lead to a smaller-than-expected coupling of high momentum particles to the small system plasma medium.  This reduction in coupling would naturally lead to a smaller-than-expected energy loss for these high momentum particles compared to propagation in larger systems at the same temperature.  This reduction in energy loss would provide a natural explanation for the current lack of evidence for high momentum particle suppression in small systems \cite{Kolbe:2015rvk}.

The finite-size effects investigated in the present study are likely to have implications in the phenomenological studies of relativistic heavy-ion collisions~\cite{Song:2010mg,Gale:2013da,Weller:2017tsr}.
A direct implication of our work is concerned with the finite-volume effect in the hot medium created by the heavy-ion collisions.
Our results suggest that the effects of such anisotropic finite volumes would not strongly affect the thermodynamics of the medium, provided that our results obtained with the PBC are directly applicable to heavy-ion physics.
The medium created in heavy-ion collisions indeed has a finite-volume and a strong anisotropic geometry. It would also be an interesting subject to pursue the connection of our study with systems having strong pressure anisotropy, such as the initial stage of the collisions.

Although we constrained our attention to a system with PBC for one direction in the present study, it is a straightforward extension of this study to perform similar analyses with other boundary conditions (see Ref.~\cite{Mogliacci:2018oea} for more details on the possible relevance of different boundary conditions). For example, it is also possible to impose anti-periodic or Dirichlet boundary conditions, instead of the PBC.
Furthermore, it is possible to impose boundary conditions for two or all the directions~\cite{Mogliacci:2018oea}. Among them, the simulation with the anti-periodic boundary conditions seems especially interesting, because the numerical analysis with this
conditions can be carried out straightforwardly, and this boundary condition eliminates the zero mode contribution (in fact, much like the Dirichlet condition~\cite{Mogliacci:2018oea}) which is the origin of the infrared divergences plaguing all theories with massless bosonic fields.

Finally, although in the present study we focused on the pressure anisotropy induced by the periodic boundary conditions in the SU(3) YM theory at finite temperature, our numerical analysis can be used for more general systems having anisotropy such as full QCD with strong magnetic field~\cite{Buividovich:2008wf,DElia:2010abb,Bali:2011qj,Bruckmann:2013oba}.

\begin{acknowledgments}
M.~K.\ thanks members of the FlowQCD and WHOT-QCD Collaborations, especially T.~Iritani, H.~Suzuki, and R.~Yanagihara for valuable discussions. Numerical simulations of this study were carried out using OCTOPUS at the Cybermedia Center, Osaka University and Reedbush at the Information Technology Center, The University of Tokyo. This work was supported by JSPS KAKENHI Grant Numbers JP17K05442, the Claude Leon Foundation, as well as the South African National Research Foundation.
\end{acknowledgments}

\bibliography{CasimirRefs}
\bibliographystyle{apsrev4-1}

\end{document}